\documentclass[12pt]{iopart}
\usepackage{graphicx}
\usepackage{dcolumn}
\usepackage{bm}
\usepackage{epstopdf}
\usepackage{textcomp}
\usepackage{iopams}  
\usepackage{float}
\usepackage{setstack}
\usepackage{multirow}
\begin{document}

\title{X-ray photoemission and resistivity studies of the Pd-covered Ce thin films}

\author{P. Skokowski$^1$, A. Marczy{\'n}ska$^1$, S. Pacanowski$^1$, T. Toli{\'n}ski$^1$, B. Szyma{\'n}ski$^1$, {\L}. Majchrzycki$^2$ and L. Smardz$^1$}
\address{$^1$ Institute of Molecular Physics, Polish Academy of Sciences\\
Smoluchowskiego 17, 60-179 Pozna{\'n}, Poland}
\address{$^2$Adam Mickiewicz University, Wielkopolska Centre for Advanced Technologies\\ Umultowska 89c, 61-614 Pozna{\'n}, Poland}
\ead{przemyslaw.skokowski@ifmpan.poznan.pl}

\begin{abstract}
We have fabricated Ce/Pd bilayers grown on the SiO$_2$ substrate by magnetron sputtering under ultrahigh vacuum. Usage of palladium layer on top of the Ce film appears to prevent effectively the sample oxidation. The thickness of the cerium films is between $10-200\ {\rm nm}$ and $10\ {\rm nm}-$thick palladium overlayers are always used. We have performed $in-situ$ X-ray photoelectron spectroscopy on the as-deposited films and $ex-situ$ electrical resistivity measurements and XRD characterization. XPS confirms a very good quality of the samples. The analysis of the Ce $3d$ spectrum suggests that the $f$ states of Ce are on the border between the fluctuating valence and localization. The resistivity measurements reveal a competition of the Kondo scattering, semiconducting and metallic behaviors as well as the influence of the dimensional effect.
\end{abstract}

\vspace{2pc}



\section{Introduction}

In recent years many research publications have been focused on the use of lanthanides \cite{ershov2013deposition,canovic2013oxidation} as protective coatings. Rare earth elements, such as Y, Ce, La, and Hf are widely known to affect the growth mechanisms of different oxides. Many surface treatments, such as sol-gel \cite{brusciotti2010characterization}, chemical conversion coating \cite{campestrini2004formation} and physical vapour deposition technique \cite{dominguez2010xps}, based on the use of cerium and its compounds have been investigated due to their low toxicity \cite{bethencourt1998lanthanide}. Recently, Canovic et al. \cite{canovic2013oxidation} reported on high-temperature oxidation behaviour of a commercial Fe-22Cr steel coated by 10 nm   Ce, 640 nm $-$ Co, and 10 nm $-$ Ce/640 nm $-$ Co thin films. They found, that applying a 640 nm Co layer on top of the 10 nm Ce layer, effectively reduces Cr evaporation and slows down the rate of alloy oxidation. However, to date there are not many literature reports on properties of single Ce layer. 

Cerium is an anomalous lanthanide exhibiting unstable $f$ occupancy, which can lead to a fluctuating valence state \cite{lawrence1981valence,gunnarsson1983electron,fuggle1983electronic}, Kondo interactions \cite{coqblin1993selected,hewson1997kondo} or a localized magnetic moment up to $2.14\ \mu_{\rm B}$ \cite{buschow_handbook_1993,gschneidner2011handbook,maple2005strongly}. Such behaviors have been widely observed in many compounds and alloys \cite{lawrence1981valence,gunnarsson1983electron,fuggle1983electronic,coqblin1993selected,hewson1997kondo,gschneidner2011handbook,maple2005strongly,strydom2006thermal,bauer2004crystalline,synoradzki2016spin,tolinski2007structural,tolinski2002electronic,tolinski2010thermoelectric}. On the other hand, the effect of dimensionality can develop transition between various structures of Ce \cite{aoki1996magnetic}, i.e. the fcc $\alpha$ (enhanced Pauli paramagnet) and the Curie$-$Weiss paramagnets dhcp $\beta$ and fcc $\gamma$ with localized $4f$ magnetic moments. The fabrication and characterization of pure Ce thin films is a challenge by the reason of its high reactivity favouring creation of the Ce oxides. Aoki $et\ al.$ \cite{aoki1996magnetic} prepared the Ce/Ta multilayers with 
$5\ {\rm nm}$ thick Ta cover layer to reduce the oxidation process. For $d_{\rm Ce} = 10\ {\rm nm}$ they found that the Ce $\alpha$ phase was stabilized at low temperatures, whereas for $d_{\rm Ce} < 1.5\ {\rm nm}$ a magnetic Ce was observed. Moreover, a crystalline state was found for the former and amorphous for the latter case. In the present research we have performed $in-situ$ X-ray photoelectron spectroscopy (XPS) on the as-deposited single Ce thin films and $ex-situ$ electrical resistivity measurements for Ce films covered with a Pd overlayer. 
\section{Experimental}
The Ce thin films were prepared at room temperature using an ultra high vacuum (UHV) magnetron sputtering \cite{smardz2005structure,smardz2000structure,smardz2002exchange}. As a substrate we have used Si(100) wafers with an oxidized surface to prevent a silicide formation \cite{skoryna2016correlation}. Therefore, we have applied a special heat treatment in UHV before deposition, in order to obtain the SiO$_2$ surface layer \cite{smardz1992oxidation}. The Pd$-$layer was deposited using a DC source. For preparation of the Ce layer the RF source was used. The thickness of the cerium films has been varied in the range $10-200\ {\rm nm}$. Typical sputtering conditions used in deposition of the cerium and palladium thin films are listed in Table 1. 

The chemical composition and the cleanness of all layers were checked $in-situ$, immediately after deposition, transferring the samples to an UHV $(4\times10^{-11}\ {\rm mbar})$ analysis chamber equipped with X-ray Photoelectron Spectroscopy (XPS), Auger Electron Spectroscopy (AES) and ion gun etching system \cite{smardz2012xps,smardz2008xps,skoryna2016xps}. The XPS spectra were measured at room temperature using a SPECS EA 10 PLUS energy spectrometer with ${\rm Al}-{\rm K}\alpha$ of 1486.6 eV. All emission spectra were measured immediately after the sample transfer in vacuum of $8\times10^{-11}\ {\rm mbar}$. The measurements were conducted following routine backing procedures $({\rm T} = 440\ {\rm K})$ of the analysis chamber, which made possible reaching a base vacuum of $4\times10^{-11}\ {\rm mbar}$. Calibration of the spectra was performed according to Baer $et\ al.$ \cite{baer1975monochromatized}. The $4f_{7/2}$ peak of gold was situated at 84.0 eV and the Fermi level was located at ${\rm E_B} = 0\ {\rm eV}$. Details of the XPS measurements can be found in \cite{smardz2012xps,smardz2008xps,skoryna2016xps}. 

The structural properties have been verified by standard $\theta-2\theta$ X-ray diffraction (XRD) using ${\rm Cu}-{\rm K}\alpha$ radiation. The surface microstructure and roughness have been studied by Atomic Force Microscopy (AFM) using tapping mode. The resistivity measurements as a function of temperature were performed on the Quantum Design Physical Property Measurement System (PPMS) in Pozna{\'n}. The usual ${\rm four}-{\rm probe}$ method has been used in the electrical measurements. All the Ce thin films prepared for the resistivity and XRD measurements before transfer from UHV to environmental conditions were covered by the Pd protective layer of about 10 and 20 nm, respectively.

\begin{table}
\caption{\label{tab1} Typical sputtering conditions.}
\centering
\begin{indented}
\item[]\begin{tabular}{@{}p{4cm}cc}
\br
Parameter & Ce & Pd\\
\mr
Rest gas pressure & \multicolumn{2}{c} {$4\times10^{-11}\ {\rm mbar}$} \\
Argon partial pressure & \multicolumn{2}{c} {$1\times10^{-3}\ {\rm mbar}$} \\
Argon purity & \multicolumn{2}{c} {99.9998$\%$} \\
Target diameter & \multicolumn{2}{c} {51.5 mm} \\
Target purity & \multicolumn{2}{c} {99$\%$}  \\
Distance between&\multicolumn{2}{c}{\multirow{2}{*}{220 mm}}\\
substrate and target&\multicolumn{2}{c}{}\\
Magnetron sputtering&\multirow{2}{*}{RF}&\multirow{2}{*}{DC}\\
method&{}\\
Sputtering power & 40 W & 35 W \\
Deposition rate & $\sim 0.04$ nm$/$s$^2$ & $\sim 0.1$ nm$/$s$^2$ \\
Substrate temperature  &\multicolumn{2}{c} {295 K}\\
\br
\end{tabular}
\end{indented}
\end{table}

\section{Results}

\begin{figure}[h]
\centering
\includegraphics[width=8cm]{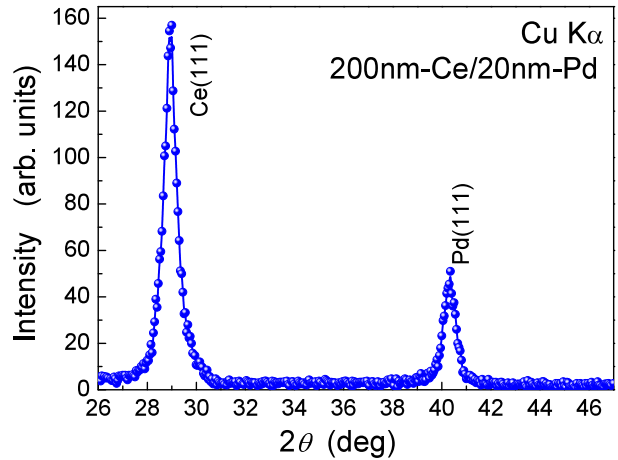}
\caption{\label{fig1}X-ray diffraction pattern for $200\ {\rm nm}-{\rm Ce}/20\ {\rm nm}-{\rm Pd}$ at RT.}
\end{figure}

As indicated in previous section to verify the structure and morphology of the studied thin films characterization by XRD and AFM techniques has been performed. \Fref{fig1} shows X-ray diffraction pattern for the sample $200\ {\rm nm}-{\rm Ce}/20\ {\rm nm}-{\rm Pd}$ measured at $300\ {\rm K}$. The Bragg reflections are well resolved both for the positions expected for Ce and Pd sublayers. The peak position of Ce thin film is shifted relative to the position expected for bulk gama fcc cerium. 

\begin{figure}[h]
\centering
\includegraphics[width=8cm]{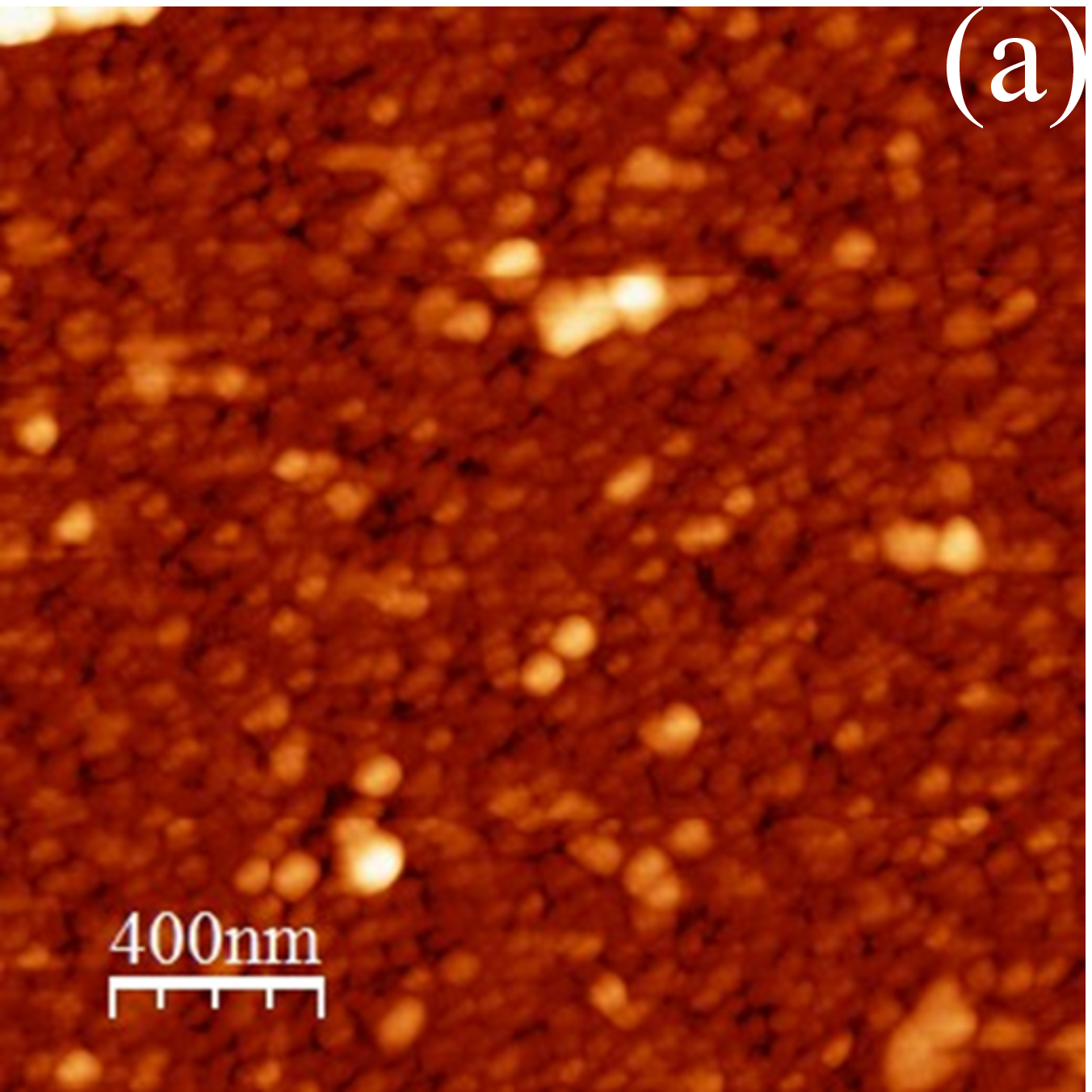}\\
\vspace{0.1cm}
\includegraphics[width=8cm]{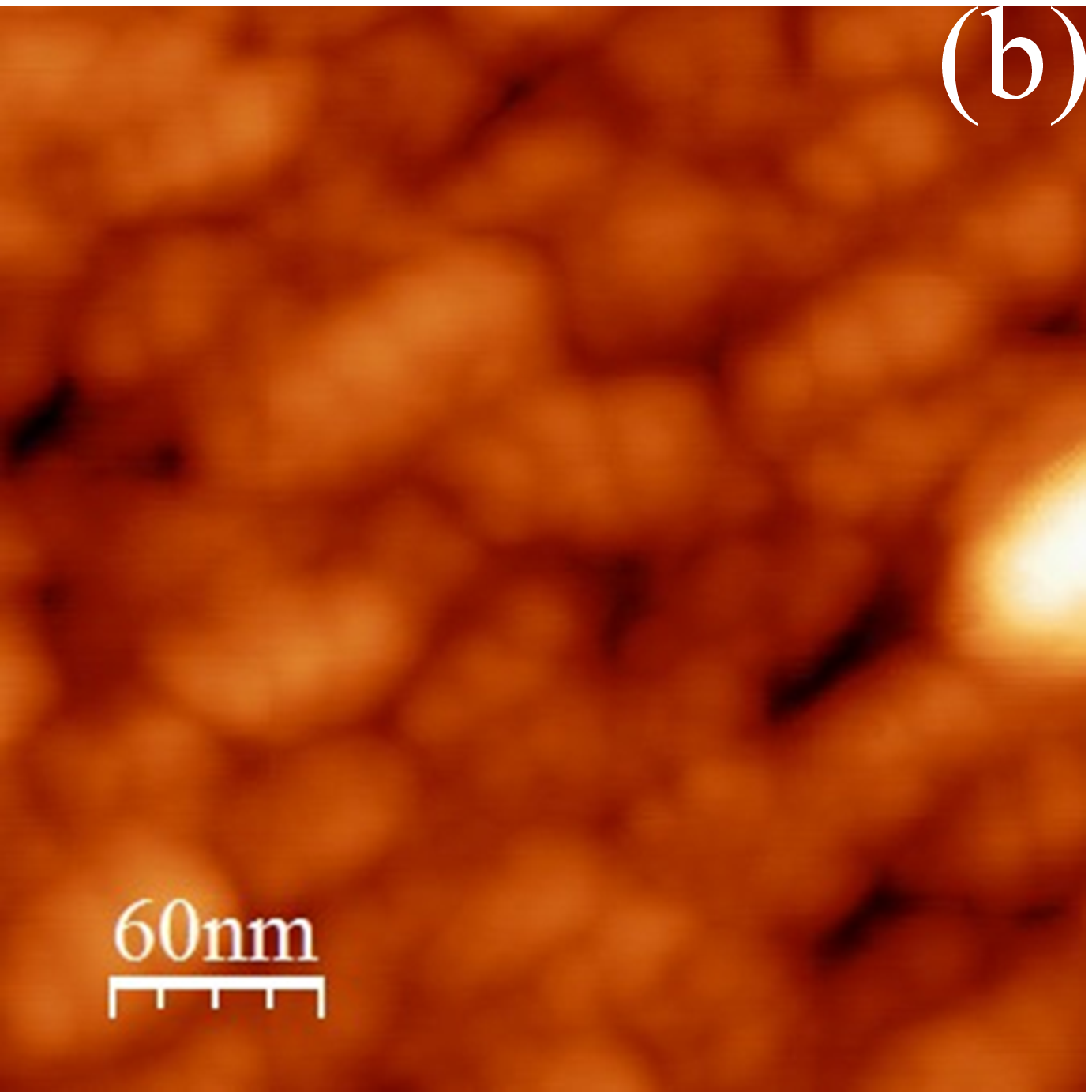}

\caption{\label{fig2} Atomic Force Microscopy images for the sample $75\ {\rm nm}-{\rm Ce}/10\ {\rm nm}-{\rm Pd}$ measured at RT. Estimated roughness parameters from image (a): ${\rm RRMS} \approx 3.24\ {\rm nm}$ and ${\rm Ra} \approx 2.10\ {\rm nm}$.}
\end{figure}

The exemplary surface morphology is illustrated in \Fref{fig2} for the sample $75\ {\rm nm}-{\rm Ce}/10\ {\rm nm}-{\rm Pd}$ at RT. It results from \Fref{fig2}a that the surface is homogeneous and at the increased resolution used in \Fref{fig2}b it can be noticed that the grains are of the lateral dimensions of about $60\ {\rm nm}$. The estimated roughness parameters from the area 2000 nm $\times$ 2000 nm (\Fref{fig2}a) were of about ${\rm RRMS} \approx 3.24\ {\rm nm}$ and ${\rm Ra} \approx 2.10\ {\rm nm}$, confirming a good quality of the films.

\begin{figure}[h]
\centering
\includegraphics[width=8cm]{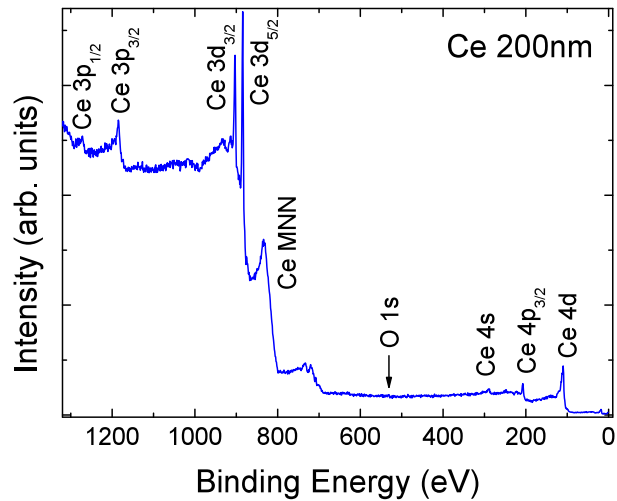}
\caption{\label{fig3} XPS (${\rm Al}-{\rm K}\alpha$) spectrum of the $d_{\rm Ce} = 200\ {\rm nm}$ film measured $in-situ$ for the $0-1400$ eV binding energy range.}
\end{figure}

Further characterization of the films concerns the electronic properties, hence, the X-ray photoemission spectroscopy has been employed as it allows to verify the chemical purity of the material in respect to the nominal composition. The survey of the entire XPS spectra have shown neither the contamination of the oxygen nor carbon confirming a very high purity of the samples. A detailed identification of the main peaks in the binding energy range $0-1400$ eV is presented in \Fref{fig3}. Due to well known high reactivity of cerium with oxygen we have prepared the thin film samples after an additional heating of the sample holder and the new substrate at $700\ {\rm K}$ for $3\ {\rm h}$ and cooling down to $295\ {\rm K}$. It appeared that after such an outgassing procedure, it is possible to prepare oxygen and carbon free Ce surface. If the outgassing procedure is too short, the sample holder before deposition will not be perfectly clean. Moreover, if the XPS measurements were performed after a few hours delay the sample surface would already be partly oxidized despite the vacuum of $4\times10^{-11}\ {\rm mbar}$ in the analysis chamber. Then the intensity of the XPS signal decreases and the valence band can be artificially broadened. For our procedure the oxygen and other surface impurities are practically absent on the Ce thin films immediately after deposition. As can be seen in \Fref{fig3}, practically no XPS signal from potential contamination of atoms like O$-$1s and C$-$1s is observed.

The analysis of the Ce $3d$ spectrum has been carried out in frames of the Gunnarsson and Sch{\"o}nhammer \cite{gunnarsson1983electron} and Fuggle $et\ al.$ \cite{fuggle1983electronic} model. The Tougaard method was used to subtract the background. The experimental spectra were fitted using a mixture of Gaussian and Lorentzian shapes.

Apart from the main peaks Ce $3d$ spectrum reveals satellites generated by the hybridization of the Ce $4f$ states with the conduction electrons. The $f^0$ and $f^2$ final states satellites has intensity sensitive to the occupancy of the $f$ states, $n_f$, and the hybridization energy, $\Delta$ , respectively. 

\begin{figure}[h]
\centering
\includegraphics[width=8cm]{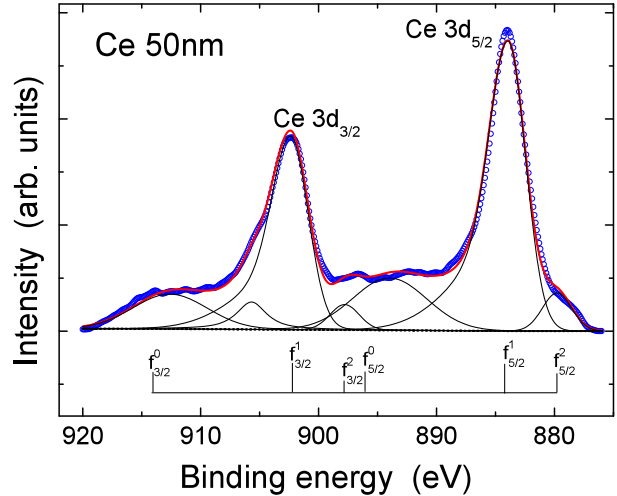}
\caption{\label{fig4} Decomposition of the XPS Ce $3d$ spectrum according to the model described in the text. The fitting line and its components are also displayed.}
\end{figure}

The intensity ratios: 
\begin{eqnarray}
r_0=\frac{I(f^0)}{I(f^0)+I(f^1)+I(f^2)}\\
r_2=\frac{I(f^2)}{I(f^1)+I(f^2)}
\end{eqnarray}
enable determination of $\Delta$ and $n_f$ from the theoretical $r_2(\Delta)$ and $r_1(n_f)$ dependences \cite{gunnarsson1983electron,fuggle1983electronic}. The decomposition presented in \Fref{fig4} revealed a very small value of the hybridization energy $\Delta = 32\ {\rm meV}$ and a slightly reduced occupancy of the $f$ states $n_f = 0.80$, which suggests that the $f$ states of Ce are on the border between a fluctuating valence and localized state. 

\begin{figure}[h]
\centering
\includegraphics[width=8cm]{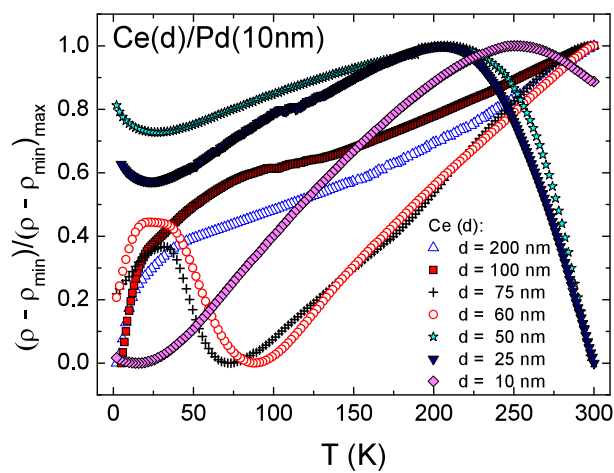}
\caption{\label{fig5} Normalized resistivity as a function of temperature for various cerium thicknesses.}
\end{figure}

In \Fref{fig5} we have presented $\rho(T)$ for the Ce(d)/Pd(10nm) bilayers with $d = 10,\ 25,\ 60,\ 75,\ 100$, and $200\ {\rm nm}$. The residual resistivity changes in the range $40-250\ \mu\Omega\ {\rm cm}$, therefore in \Fref{fig5} normalized results are shown to enable a comparison of different curves. It is evident that a competition of various contributions is responsible for the significant changes of the $\rho(T)$ behavior while changing the Ce film thickness. We interpret the visible anomalies as resulting from the Kondo scattering due to the Kondo impurity effect within the Ce/Pd interface, the Crystalline Electric Field (CEF) effect at the upper range of the temperatures studied, and finally the dimensional effect appearing when the thickness of the bilayer becomes comparable with the scattering length. The behavior visible in \Fref{fig5} leads to the conclusion that the dimensional effect switches on below $d_{\rm Ce} = 60\ {\rm nm}$.

\begin{figure}[h]
\centering
\includegraphics[width=8cm]{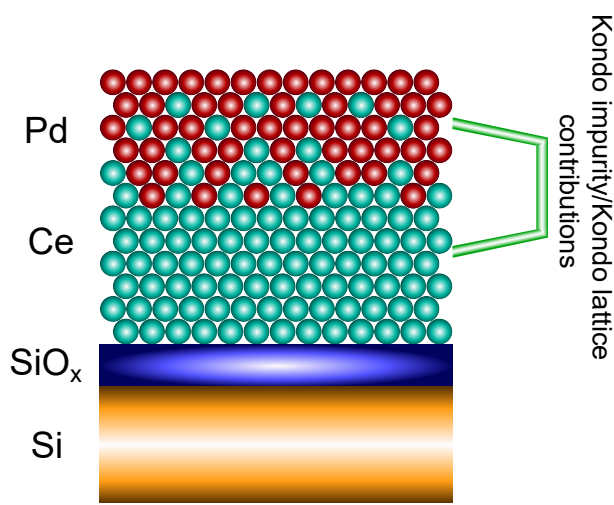}
\caption{\label{fig6} Illustration of the Ce(d)/Pd(10nm) bilayers on the Si substrate.}
\end{figure}

A detailed scenario can be proposed based on the illustration displayed in \Fref{fig6}. It is well known that the interface between Pd and Ce is relatively broad due to an interface mixing bulk Ce and Pd form Ce-Pd alloys [x]  ([x] J.R. Thomson, J. Less Comon Metals 13 (1967) 307. In our case, this alloy formation is not homogenous and only near Pd-Ce interface takes place. Therefore, such an ultrathin layer could be not visable in XPS and XRD experiment. For thick Ce films $(60-200\ {\rm nm})$ a metallic resistivity dominates above $100\ {\rm K}$, but at lower temperatures the dilution within Ce/Pd interface provides a Kondo and Kondo lattice contribution to the scattering, which results in the appearance of peaks at about $25\ {\rm K}$ implying a presence of correlations. These peaks are present as bumps for $d_{\rm Ce} = 200\ {\rm nm}$ and $100\ {\rm nm}$ and are better resolved below $100\ {\rm nm}$, when the contribution of metallic conductivity is reduced due to the size effect. For $d \leq 50\ {\rm nm}$ a semiconductor-like resistivity is observed at high temperatures and a high temperature peak develops at $200\ {\rm K}$ due to increasing metallic conductivity at lower temperatures. The low temperature peak ascribed previously to Kondo ${\rm lattice}-{\rm like}$ behavior switches to a minimum caused by the Kondo impurity mechanism. The last possibility is probable as, due to the ultrathin Ce film, there is a dropping amount of Ce ions within the interfaces and the ions start to behave like uncorrelated impurities.

\section{Conclusion}
A good quality Ce films covered with Pd overlayer have been fabricated, as confirmed by XRD and AFM studies of the structural properties and the morphology, respectively. The X-ray photoemission has shown that $10\ {\rm nm}$ thick Pd layer effectively prevents the Ce film oxidation. The analysis of the Ce $3d$ spectrum using the Gunnarsson$-$Sch{\"o}nhammer$-$Fuggle model indicates that the Ce $f$ states are on the border between the fluctuating valence and localization. A reasonable explanation of the temperature dependence of the resistance of the Ce films with thickness varied in the range $10-200\ {\rm nm}$ is possible if one assumes importance of the mixing within the Ce/Pd interface. Then, depending on the Ce film thickness, Kondo ${\rm impurity}-{\rm like}$ or Kondo ${\rm lattice}-{\rm like}$ behavior can be concluded. At the upper range of the temperatures studied the Crystalline Electric Field influence is also possible. The dimensional effect expected when the thickness becomes comparable with the scattering length switches on below $d_{\rm Ce} = 60\ {\rm nm}$.


\section*{References}
\begin{thebibliography}{10}
\expandafter\ifx\csname url\endcsname\relax
  \def\url#1{{\tt #1}}\fi
\expandafter\ifx\csname urlprefix\endcsname\relax\def\urlprefix{URL }\fi
\providecommand{\eprint}[2][]{\url{#2}}

\bibitem{ershov2013deposition}
Ershov S, Druart M~E, Poelman M, Cossement D, Snyders R and Olivier M~G 2013
  {\em Corros. Sci.\/} {\bf 75} 158--168

\bibitem{canovic2013oxidation}
Canovic S, Froitzheim J, Sachitanand R, Nikumaa M, Halvarsson M, Johansson L~G
  and Svensson J~E 2013 {\em Surf. Coat. Technol.\/} {\bf 215} 62--74

\bibitem{brusciotti2010characterization}
Brusciotti F, Batan A, De~Graeve I, Wenkin M, Biessemans M, Willem R, Reniers
  F, Pireaux J, Piens M, Vereecken J {\em et~al.\/} 2010 {\em Surf. Coat.
  Technol.\/} {\bf 205} 603--613

\bibitem{campestrini2004formation}
Campestrini P, Terryn H, Hovestad A and De~Wit J 2004 {\em Surf. Coat.
  Technol.\/} {\bf 176} 365--381

\bibitem{dominguez2010xps}
Dominguez-Crespo M~A, Torres-Huerta A~M, Rodil S, Brachetti-Sibaja S,
  De~La~Cruz W and Flores-Vela A 2010 {\em J. Appl. Elektrochem.\/} {\bf 40}
  639--651

\bibitem{bethencourt1998lanthanide}
Bethencourt M, Botana F, Calvino J, Marcos M and Rodriguez-Chacon M 1998 {\em
  Corros. Sci.\/} {\bf 40} 1803--1819

\bibitem{lawrence1981valence}
Lawrence J, Riseborough P and Parks R 1981 {\em Rep. Prog. Phys.\/} {\bf 44} 1

\bibitem{gunnarsson1983electron}
Gunnarsson O and Sch{\"o}nhammer K 1983 {\em Phys. Rev. B\/} {\bf 28} 4315

\bibitem{fuggle1983electronic}
Fuggle J, Hillebrecht F~U, Zo{\l}nierek Z, L{\"a}sser R, Freiburg C, Gunnarsson
  O and Sch{\"o}nhammer K 1983 {\em Phys. Rev. B\/} {\bf 27} 7330

\bibitem{coqblin1993selected}
Coqblin B, Arispe J, Bhattacharjee A and Evans S 1993 Kondo effect and heavy
  fermions {\em Frontiers in Solid State Sciences, Selected Topics in
  Magnetism\/} vol~2 ed Gupta L~C and Multani M~S (Singapore: World Scientific)
  p~75

\bibitem{hewson1997kondo}
Hewson A~C 1997 {\em Cambridge Studies in Magnetism\/} vol~2 {\em The Kondo
  problem to heavy fermions\/} (Cambridge: Cambridge University Press)

\bibitem{buschow_handbook_1993}
Buschow K~H~J 1993 {\em Handbook of magnetic materials\/} vol~7 (Amsterdam:
  Elsevier North Holland)

\bibitem{gschneidner2011handbook}
Thalmeier P and Zwicknagl G 2004 Unconventional {Superconductivity} and
  {Magnetism} in {Lanthanide} and {Actinide} {Intermetallic} {Compounds} {\em
  Handbook on the physics and chemistry of rare earths\/} vol~34 ed Gschneidner
  K~A, B{\"u}nzli J~C~G and Pecharsky V~K (Amsterdam: Elsevier North Holland)

\bibitem{maple2005strongly}
Maple M~B 2005 {\em J. Phys. Soc. Jpn.\/} {\bf 74} 222--238

\bibitem{strydom2006thermal}
Strydom A, Paschen S and Steglich F 2006 {\em Physica B\/} {\bf 378} 793--794

\bibitem{bauer2004crystalline}
Bauer E, Christianson A, Lawrence J, Goremychkin E, Moreno N, Curro N, Trouw F,
  Sarrao J, Thompson J, McQueeney R {\em et~al.\/} 2004 {\em J. Appl. Phys.\/}
  {\bf 95} 7201--7203

\bibitem{synoradzki2016spin}
Synoradzki K and Toli{\'n}ski T 2016 {\em Mater. Chem. Phys.\/} {\bf 177}
  242--249

\bibitem{tolinski2007structural}
Toli{\'n}ski T 2007 {\em Mod. Phys. Lett. B\/} {\bf 21} 431--454

\bibitem{tolinski2002electronic}
Toli{\'n}ski T, Pugaczowa-Michalska M, Che{\l}kowska G, Szlaferek A and
  Kowalczyk A 2002 {\em Phys. Status Solidi B\/} {\bf 231} 446--450

\bibitem{tolinski2010thermoelectric}
Toli{\'n}ski T, Zlati{\'c} V and Kowalczyk A 2010 {\em J. Alloy. Compd.\/} {\bf
  490} 15--18

\bibitem{aoki1996magnetic}
Aoki Y, Sato H, Komaba Y, Kobayashi Y, Sugawara H, Hashimoto S, Yokoyama T and
  Hanyu T 1996 {\em Phys. Rev. B\/} {\bf 54} 12172

\bibitem{smardz2005structure}
Smardz L 2005 {\em J. Alloy. Compd.\/} {\bf 395} 17--22

\bibitem{smardz2000structure}
Smardz L, Smardz K and Niedoba H 2000 {\em J. Magn. Magn. Mater.\/} {\bf 220}
  175--182

\bibitem{smardz2002exchange}
Smardz L 2002 {\em J. Magn. Magn. Mater.\/} {\bf 240} 273--276

\bibitem{skoryna2016correlation}
Skoryna J, Wachowiak M, Marczy{\'n}ska A, Rogowska A, Koczorowski W, Czajka R
  and Smardz L 2016 {\em Surf. Coat. Technol.\/} {\bf 303} 119--124

\bibitem{smardz1992oxidation}
Smardz L, K{\"o}bler U and Zinn W 1992 {\em J. Appl. Phys.\/} {\bf 71}
  5199--5204

\bibitem{smardz2012xps}
Smardz L, Nowak M and Jurczyk M 2012 {\em Int. J. Hydrogen Energ.\/} {\bf 37}
  3659--3664

\bibitem{smardz2008xps}
Smardz K, Smardz L, Okonska I, Nowak M and Jurczyk M 2008 {\em Int. J. Hydrogen
  Energ.\/} {\bf 33} 387--392

\bibitem{skoryna2016xps}
Skoryna J, Pacanowski S, Marczy{\'n}ska A, Werwi{\'n}ski M, Rogowska A,
  Wachowiak M, Czajka R, Smardz L {\em et~al.\/} 2016 {\em Surf. Coat.
  Technol.\/} {\bf 303} 125--130

\bibitem{baer1975monochromatized}
Baer Y, Busch G and Cohn P 1975 {\em Rev. Sci. Instrum.\/} {\bf 46} 466--469

\end{thebibliography}
\end{document}